\documentclass[pra,twocolumn,showpacs]{revtex4}
\usepackage{epsfig}
\usepackage{amsbsy}
\usepackage{amsmath}
\usepackage{graphicx}
\usepackage{latexsym}
\usepackage{amssymb}
\usepackage{color}
\usepackage[normalem]{ulem}

\begin{document}

\title{Coherent tunneling of collective excitation of Bose-Einstein condensate in a double-well potential}
\author{Yaojun Ying, Haoyu Wang, Ming Zhang and Haibin Li}

\affiliation{School of Physics, Zhejiang
University of Technology, Hangzhou 310023, China.
}

\date{\today}

\begin{abstract}
The Josephson effect can be observed in a Bose-Einstein condensate in a double-well potential, which is attributed to the tunneling of bosons between two wells. We propose a multi-mode theory to investigate the dynamics of local excitations in a one-dimensional condensate in a double-well potential. We show that the system can be described by two independent two-mode models. The Josephson oscillation and the self-trapping of local collective excitations are predicted analytically and confirmed by numerical simulation.

\end{abstract}


\maketitle
\section{\label{sec:level 1}Introduction}
The Josephson effect was first predicted in 1962~\cite{Josephson}, and later verified experimentally~\cite{Shap}. It is one of the typical manifestations of macroscopic quantum coherence between two superconductors weakly coupled by a thin insulating layer. Due to such coherence, Cooper pairs can tunnel through the insulator. The Josephson effect can be attributed to couplings caused by overlapping of wave functions. The Josephson effect can also be realized in other scenarios, such as superfluidity~\cite{Perev,Backh,Davis,Sukna}, Bose-Einstein condensate (BEC)~\cite{Catal,Albiez,Levy}, and excitation-polarization in microcavities~\cite{Abbar}.

A great deal of research has been carried out on the problem of a BEC in a double-well potential, both theoretically~\cite{Juha,Milburn,Smerzi,Zapata,Raghavan,Giovan,Sake,Anan,Giovan2,Ichih,Juli,Jezek,Burchi,Marti} and experimentally~\cite{Catal,Albiez,Levy,Leb,Trenk}, focusing on the Bose-Josephson junction. Interatomic interactions are found to significantly affect Bose-Josephson dynamics, leading to phenomena such as condensate self-trapping~\cite{Milburn,Smerzi,Albiez}.

Strictly solving the dynamics of a bosonic condensate confined in a potential well requires starting from the basic principles of quantum mechanics and solving the Schr\"{o}dinger equation. However, due to the presence of interactions, it is a typical quantum many-body problem that cannot be solved, in general. Considering the weakness of interactions between atoms in the condensate, the dynamics of the system can be treated with mean-field theory, leading to the Gross-Pitaevskii (GP) equation. The two-mode approximation applied to the GP equation can then describe a condensate in a double-well potential. In this scenario, the two modes often refer to the ground state and the first excited state of the global double-well potential, or alternatively, the two local ground states if each well is treated independently~\cite{Milburn}.

As is well known, quantum tunneling is a key manifestation of quantum mechanics. In the non-interacting case, the dynamics of the condensate in a double-well potential reduces to a single-particle tunneling between the traps. In the actual condensate, interactions cannot be ignored. The tunneling of the condensate can no longer be regarded as the behavior of a single particle. However, in the framework of the mean-field approach, previous studies of condensates in double‑well potentials — both experimental and theoretical — have primarily adopted a statistical viewpoint, focusing on quantities such as the particle number distribution and its imbalance between two wells. The nonlinear term in GP equation arising from the interactions between particles leads to a shift in the oscillation frequency of the population imbalance and the emergence of self-trapping~\cite{Milburn,Smerzi}. Nevertheless, at sufficiently low temperatures, the dynamics of the system is governed by collective excitations rather than single-particle ones, indicating that a purely single-particle perspective is insufficient.

In fact, collective excitation is a crucial feature of condensate dynamics and has been extensively studied both in theory and experiment~\cite{Dalfovo}. Moreover, collective excitations serve as an essential diagnostic tool for investigating finite-temperature and finite-size effects in condensates. In double-well potentials, collective excitation modes have been studied using the Bogoliubov approach~\cite{Burchi,Marti}, which yields more accurate frequencies for the population imbalance. In this paper, we will explore the dynamics of a BEC in a double-well potential when a collective excitation is generated locally in one well, e.g., a local dipole oscillation. We find that the collective excitation can also undergo the Josephson oscillations between the two wells, and when the interactions are sufficiently large, the collective excitation can become self-trapping in one well.

This paper is organized as follows. In Sec.\ref{sec:level 2}, we establish a model to investigate the dynamics of local collective excitations of BEC in a double-well potential by incorporating the first excited local mode in each well. The oscillation and self-trapping of local collective excitations are predicted. In Sec.\ref{sec:level 3}, we adopt numerical simulations and confirm our predictions of collective excitations. Finally, we draw our conclusions in Sec.\ref{sec:level 4}.

\section{\label{sec:level 2}Extended two-mode model}
For simplicity, let us consider a one-dimensional condensate system of $N$ particles trapped in a double-well potential, whose dynamics is described by the GP equation,
\begin{equation}
\label{eq:1}
i\hbar\frac{\partial }{\partial t}\psi(x,t)=(H_{0}+u_{0} \vert \psi(x,t) \vert ^2)\psi(x,t),
\end{equation}
with
\begin{equation}
H_{0}=-\frac{\hbar^2}{2m}\nabla^2+V(x),
\end{equation}
where $m$ is the mass of a particle, and $u_0=gN$, with $g$ denoting the strength of interaction in one dimension. The condensate wavefunction is normalized to unity, $\int \vert \psi \vert^2 dx=1 $. The trap potential $V(x)$ is a double-well potential.
Throughout this paper, we set $\hbar=m=\omega=1$.

\begin{figure}
\includegraphics[width=0.45\textwidth]{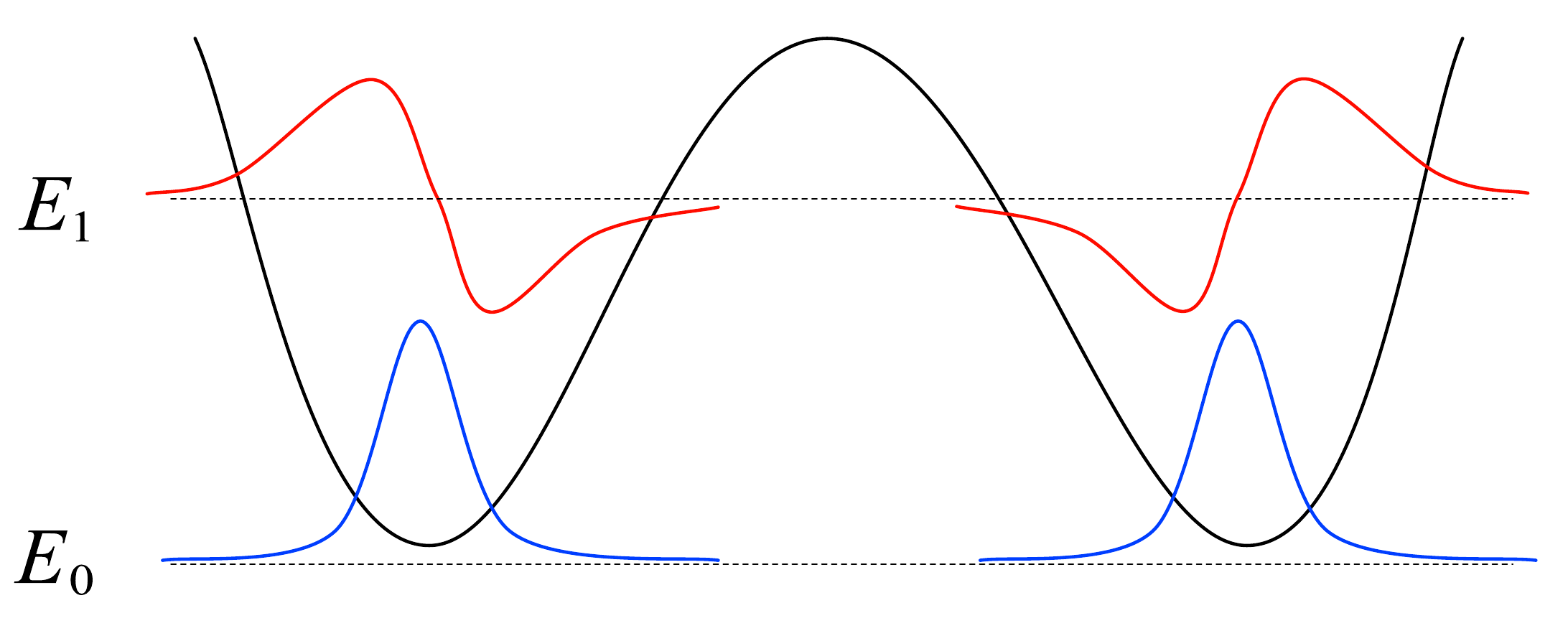}
\caption{Schematics of the symmetric double-well potential
The local ground state and first excited state are shown for each well. $E_{0}$ and $E_{1}$, are the eigenenergy of local ground state and the local first excited state.}
\label{fig:1}       % Give a unique label
\end{figure}

In the standard two-mode approximation, only two local ground states in each well or the two lowest energy eigenstates of the global potential are considered~\cite{Milburn,Smerzi}. The dynamics of the system is reflected in the population imbalance of two wells. However, in experiments or numerical simulations, the method used to generate an initial population imbalance often excites additional collective modes in the trap, which are not captured by the standard two-mode model. In order to investigate the BEC dynamics, including higher modes, in a double-well potential, we should modify the two-mode approximation to incorporate more modes. The main assumption used in the two-mode approximation for a global double-well potential is that only the two lowest modes are populated. Here, we apply this assumption to each well, assuming that the local ground state and the local first excited state are populated, as illustrated in Fig.~\ref{fig:1}. Consequently, the wavefunction of the total condensate in the global well can be written as

\begin{equation}
\label{eq:3}
\Psi(x,t)=a_{0}^{l}(t)\psi_{0}^{l}
+a_{1}^{l}(t)\psi_{1}^{l}+
a_{0}^{r}(t)\psi_{0}^{r}+
a_{1}^{r}(t)\psi_{1}^{r},
\end{equation}
where $\psi_{i}^{l(r)}(i=0,1)$ represent the local ground state mode and the local first excited mode of the condensate in the left (right) well, and satisfy the orthogonality condition, 
$(\int (\psi_{i}^{\alpha})^{*}\psi_{j}^{\beta}dx=0,\alpha,\beta=l,r\quad and \quad  i\neq j)$. 
On the other hand, for the global double-well potential, we should consider four eigenstates with the lowest energy levels, $\phi_{i}$ with $i=0,1,2,3$. Then the four local modes should be expressed as a combination of symmetric and asymmetric global eigenstates as
\begin{equation}
\label{eq:4}
\begin{split}
\psi_{0}^{l,r}=\frac{1}{2}(\phi_{0}\pm
\phi_{1})\\
\psi_{1}^{l,r}
=\frac{1}{2}(\phi_{2}\pm
\phi_{3}).
\end{split}
\end{equation}
After substituting the ansatz (\ref{eq:3}) into the GP equation (\ref{eq:1}), we obtain coupled functions of the mode amplitudes. In this process, the coupling between the local modes caused by nonlinear interaction is ignored. We further find that coupling between local modes in different wells is determined by mode symmetry: only modes of the same symmetry are coupled. Consequently, the dynamic equation of condensate decouples into two independent equations,

\begin{equation}
\label{eq:6}
\begin{split}
i\frac{da_{0}^{l}}{dt}=E_{0}a_{0}^{l}
-K_{0}a_{0}^{r}+U_{0}\vert a_{0}^{l} \vert ^2 a_{0}^{l}\\
i\frac{da_{0}^{r}}{dt}=E_{0}a_{0}^{r}
-K_{0}a_{0}^{l}+U_{0}\vert a_{0}^{r} \vert ^2 a_{0}^{r},
\end{split}
\end{equation}
and

\begin{equation}
\label{eq:7}
\begin{split}
i\frac{da_{1}^{l}}{dt}=E_{1}a_{1}^{l}
-K_{1}a_{1}^{r}+U_{1}\vert a_{1}^{l} \vert ^2 a_{1}^{l}\\
i\frac{da_{1}^{r}}{dt}=E_{1}a_{1}^{r}
-K_{1}a_{1}^{l}+U_{1}\vert a_{1}^{r} \vert ^2 a_{1}^{r}.
\end{split}
\end{equation}
where $E_{i}(i=0,1)$ are the eigenenergies of the local ground state and the local first excited state, $K_{i}=\int \psi_{i}^{l}H_{0}\psi_{i}^{r}dx,(i=0,1)$ are the coupling strengths between the two local ground states and the two first excited states. $U_{i}={u_0}\int (\psi_{i}^{l})^{4}dx=u_0\int (\psi_{i}^{r})^{4}dx,(i=0,1$), represents the effective on-site interaction energy for the local ground state and local first excited state due to nonlinear interaction within each well.

Eq.(\ref{eq:6}) and Eq.(\ref{eq:7}) share an identical mathematical structure which describes the dynamic of a two-mode model, but they represent different physics context. Eq.(\ref{eq:6}) corresponds to the standard two-mode model for the coupled dynamics of the two local ground states, whereas  Eq.(\ref{eq:7}) governs the coupled dynamics of the two local first excited states.

In Eq.(\ref{eq:6}) and Eq.(\ref{eq:7}), the mode amplitudes can be written as $a_{i}^{l,r}=\sqrt{N_{i}^{l,r}}e^{i\theta_{i}^{l,r}}(i=0,1)$
, where $N_{i}^{l,r}(i=0,1)$ and $\theta_{i}^{l,r},(i=0,1)$ denote the number of particles and phase in the local ground or first excited state in the left ($l$) or right ($r$) well. The total particle number is $N=\sum_{\alpha,i}{N_{i}^{\alpha},(\alpha=l,r, and \quad i=0,1)}$. Defining the population imbalance $z_{i}=\frac{N_{i}^{l}-N_{i}^{r}}
{N_{i}^{l}+N_{i}^{r}},(i=0,1)$ and the relative phase $\theta_{i}=\theta_{i}^{l}-\theta_{i}^{r},(i=0,1)$, we can rewrite Eq.(\ref{eq:6}) and (\ref{eq:7}) as:

\begin{equation}
\label{eq:8}
\begin{aligned}
\dot{z}_{i}&=-\sqrt{1-z_{i}^2}
\sin(\theta_{i})\\
\dot{\theta}_{i}&=A_{i}z_{i}+
\frac{z_{i}}{\sqrt{1-z_{i}^2}}\cos(\theta_{i}),
\end{aligned}
\end{equation}
where $A_{i}=(N_{i}^{l}+N_{i}^{r})
U_{i}/(4K_{i})$ ($i=0,1$). This set of equations matches the standard two-mode approximation for the condensate in a double-well potential, originally formulated for the local ground states alone. Hence, in our approach, the collective excitations of a BEC in a double-well potential can be modeled by two independent, nonrigid pendulums with Hamiltonian $H_{i}=\frac{A_{i}}{2}z_{i}^{2}-\sqrt{1-z_{i}^{2}}\cos(\theta_{i}),(i=0,1)$. As is well known, the original two-mode model exhibits Josephson oscillation and self-trapping. We anticipate that these phenomena will also appear in our two independent models,  and the dynamics of the whole system will be determined by the superposition of the dynamics of the two models.

Let us consider a simple scenario: initially, the two local ground states share the same population, each containing half of the total condensate system. In the right well, the lowest collective excitation is triggered, occupying the first excited state with a few particles, whereas in the left well, the first excited state remains unoccupied. Under these conditions, due to the weakness of the perturbation, the population imbalance of two local ground states will remain zero, exhibiting no oscillation. In contrast, the two local first excited states start with a population imbalance of unity, so a Josephson oscillation or self-trapping is expected to emerge in different dynamic regimes, which can be addressed by the particle number. Nevertheless, the overall system dynamics must account for both subsystems. The small occupation of the first excited state in the right well means that collective excitation is stimulated locally. In the parabolic approximation of local potential, this collective excitation is a dipole oscillation, which corresponds to the motion of the center of mass. When Josephson tunneling occurs for the population in the first excited state, the occupation in the left well becomes nonzero, indicating that the dipole mode has tunneled from one well to the other. In contrast, if self-trapping sets in, the dipole mode remains localized in its original well and does not tunnel to the opposite side.

\section{\label{sec:level 3}Numerical simulation}

In the following, to verify our analytical predictions, we perform numerical simulations of the system. The double-well potential is chosen as
\begin{equation}
\label{eq:2}
V(x,t)=\frac{1}{2}x^2+V_{0}e^{-2x^2/\Omega^2},
\end{equation}
where $V_{0}$ denotes the barrier height and $\Omega$ represents the barrier width. We prepare the initial state at $t=0$ as the ground state of the system. Next, we keep the condensate in the left well at rest and displace the condensate in the right well to the right by a distance $a$, which is chosen to be sufficiently small to ensure that the perturbation remains within the linear limit. Afterward, the condensate is allowed to evolve freely.
\begin{figure}
 \includegraphics[width=0.45\textwidth]{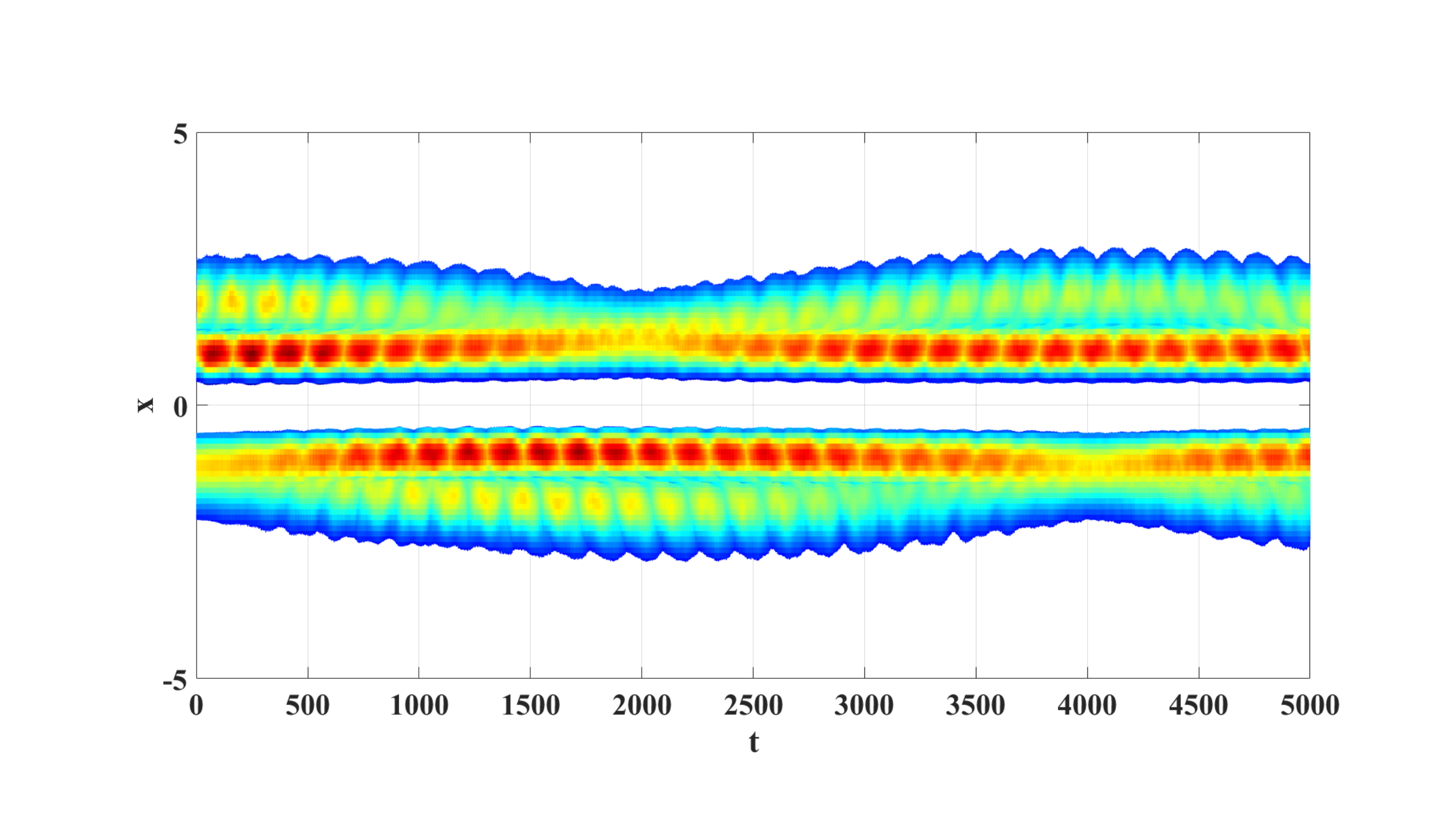}
\caption{Time evolution of the particle density of the condensate in a double-well potential. The barrier height is $V_{0}=100$ and the nonlinear interaction strength is ${u_0}=0.1$. The initial displacement parameter is $a=0.6$.}
\label{fig:2}       % Give a unique label
\end{figure}

In Fig.\ref{fig:2}, we show the time evolution of the condensate density. It can be seen that at the beginning, the initial conditions lead to a deviation of the wave function of the condensate in the right well from the ground state, which indicates that a collective excitation is generated there, with some particles occupying higher energy levels. Consequently, the condensate wavefunction becomes a superposition of the ground and excited states, leading to an oscillation of condensate. Meanwhile, the condensate in the left well essentially maintains its initial shape, showing no significant change. As time evolves, the amplitude of oscillations in the right well gradually decreases, and the density profile approaches that of the ground state. During the same time, the condensate in the left well becomes deformed, leading to oscillations around its local equilibrium position. That is to say, the local collective excitation in the right well tunnels through the barrier into the left well. In the subsequent time evolution, this local collective excitation will tunnel through the barrier again and then oscillate around the barrier. This oscillation is fundamentally different from the usual Josephson oscillation of the population imbalance between the two wells. First, under our chosen initial condition, the population imbalance is zero, which would not induce any oscillation of population imbalance in the two-mode model. However, the local collective excitation still exhibits tunneling and oscillatory behavior. We also note that the tunneling of local collective excitation will also cause a population imbalance, which is not captured by the simple two-mode theory. Secondly, in the two-mode theory describing population imbalance, no local excitations are present in either well and only the amplitudes of the two modes vary in time.

\begin{figure}
\includegraphics[width=0.45\textwidth]{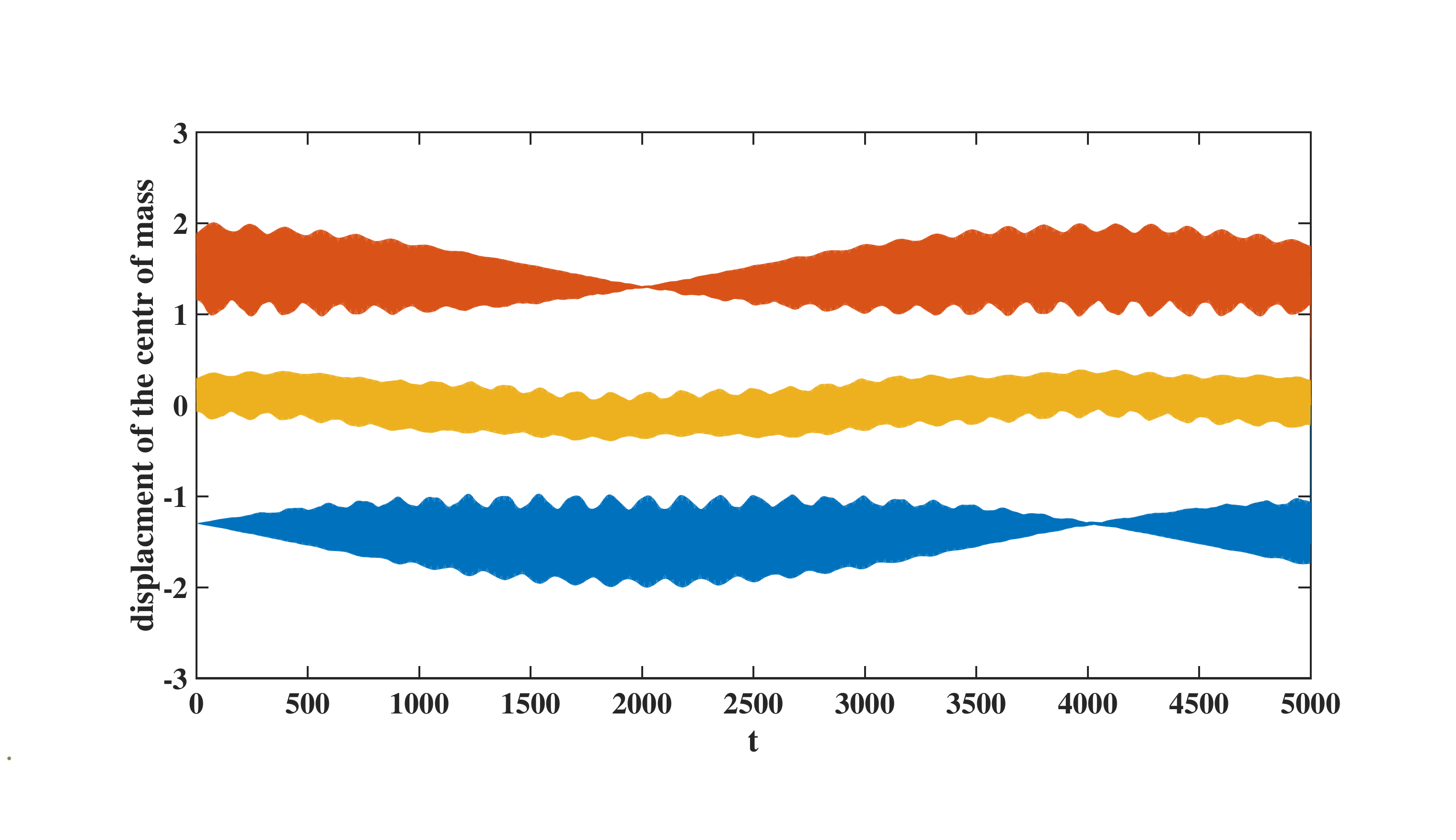}
\caption{Time evolution of the center of mass of condensate in the left well(blue line), the right well(red line), and the entire double-well system(orange line). Here, $V_{0}=100$, $u_0=0.1$ and $a=0.6$.}
\label{fig:3}       % Give a unique label
\end{figure}

In order to show the tunneling of local excitation more clearly, we calculate the time evolution of the center of mass of condensate in each well, respectively, as shown in Fig.\ref{fig:3}. We also calculate the center of mass of the entire system and plot it in Fig.\ref{fig:3}. As is well known, the oscillation of the center of mass corresponds to the lowest collective excitation, namely the dipole oscillation. One can clearly note that the collective excitation first appears in the right well, where the center of mass does periodic oscillation around the center of the right well.  Meanwhile, the center of mass in the left well stays at equilibrium point at the beginning. As time goes on, the oscillation amplitude in the right well diminishes while that in the left well increases, indicating a tunneling of the oscillation of the center of mass, i.e., the local dipole oscillation of the BEC. In previous studies of a BEC in a double-well potential via the two-mode approximation, the Josephson oscillation was understood as particle tunneling that induces a population imbalance. It should be noted that the tunneling of a collective excitation is a distinct phenomenon, which indicates more intrinsic many-body dynamics of BEC because of interaction between particles.

In population imbalance Josephson oscillations, the effect of nonlinear interactions is primarily to shift the oscillation frequency. Therefore, we also study the influence of nonlinear effects on collective excitation tunneling. We calculate the change in mode oscillation with increasing nonlinear interactions. The results indicate that as the nonlinear interaction gradually increases, the tunneling frequency also increases.

To further understand the tunneling of collective excitations, we apply our four-mode approximation to this 1D system. Recall our initial condition: the BEC in the right well is excited, whereas the BEC in the left well remains in its local ground state. As the excitation is weak, we can assume that the population of the ground state is still very close to half of the total number of particles $0.5N$. Thus, for the two-mode model constituted by the two local ground states, due to the initial zero population imbalance condition,$z_{\textcolor{red}{0}}=\frac{N_{1}^{l}-N_{1}^{r}} 
{N_{1}^{l}+N_{1}^{r}}=0$, there is no population transferred between the ground states in the two wells. On the other hand, although the first excited mode of the left well has a small population, its counterpart in the right well is zero, giving an initial relative imbalance of negative unity,$z_{\textcolor{red}{1}}=\frac{N_{2}^{l}-N_{2}^{r}}
{N_{2}^{l}+N_{2}^{r}}=-1$. This leads to a Josephson oscillation with maximum amplitude in the second two-mode model, which involves the two local first excited states. In our simulation, the initial local perturbation induces a collective excitation of the BEC in the right well. Due to tunneling, the population in the first excited state of the right well transfers to the first excited state of the left well, resulting in a local dipole oscillation there. After a period of time, this local dipole oscillation tunnels back to the right well, thereby forming a Josephson oscillation of collective excitation.
\begin{figure}
\includegraphics[width=0.45\textwidth]{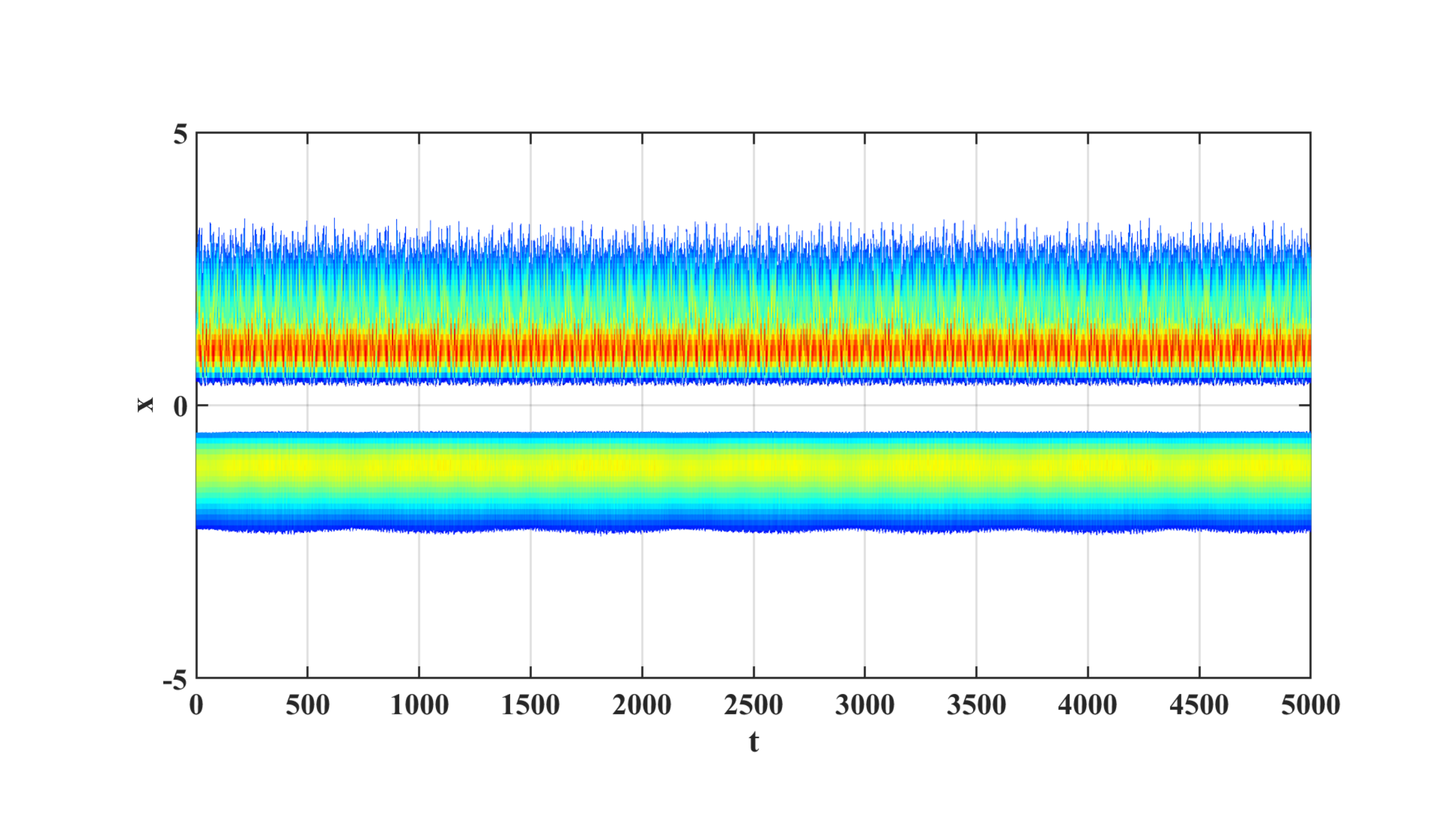}
\caption{Time evolution of the particle density of condensate in a double-well potential. The barrier height is $V_{0}=100$ and the nonlinear interaction strength is $u_0=4$. The initial displacement parameter is $a=0.6$. }
\label{fig:4}       % Give a unique label
\end{figure}

\begin{figure}
\includegraphics[width=0.45\textwidth]{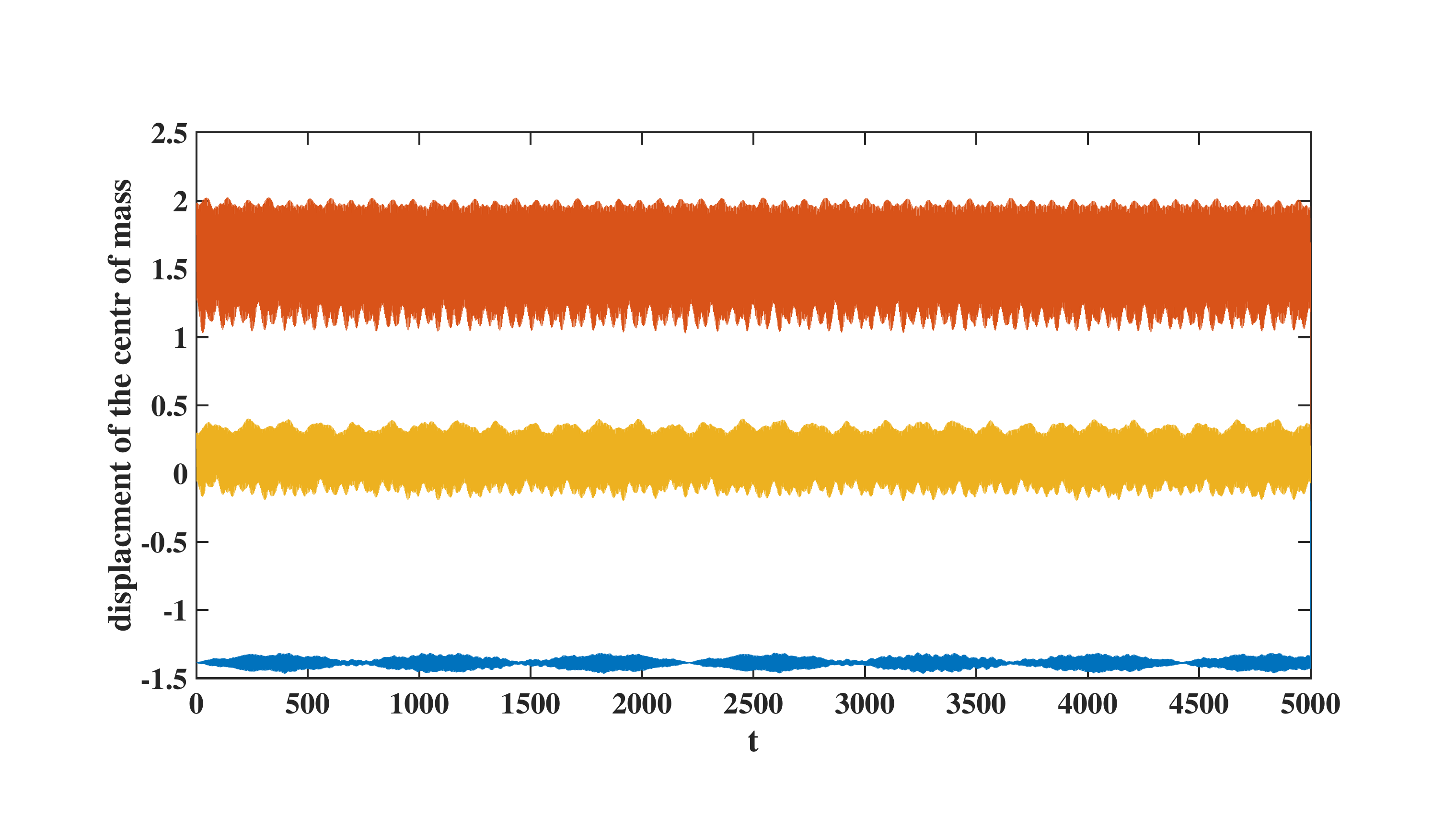}
\caption{Time evolution of the center of mass of condensate in the left well (blue line), the right well (red line),  and the entire double-well system (orange line).Here, $V_{0}=100$, $u_0=4$ and $a=0.6$.}
\label{fig:5}       % Give a unique label
\end{figure}

When the nonlinear interaction is sufficiently large (for example, $g = 4$), the condensate dynamics exhibit entirely different characteristics. As seen in Fig.~\ref{fig:4}, the time evolution of the condensate density reveals that the local collective excitation excited in the right well always remains confined to its initial well and does not tunnel to the other well. Meanwhile, the condensate in the left well remains essentially in its original state without generating collective excitations. This indicates that the collective excitation is confined to the right well. This feature appears more clearly in the time evolution of the center of mass of condensate in the left and right wells, as shown in Fig.~\ref{fig:5}. The center of mass of the BEC in the right well oscillates around its equilibrium position with a nearly constant amplitude. Meanwhile, in the left well, the center of mass oscillates with an amplitude much smaller than that in the right well. This indicates that the local collective excitation, e.g., local dipole oscillation, remains locked in its original well, which is called the self-trapping of the local excitation. We also calculate the population imbalance between the two wells in this case, finding that it oscillates but with a small amplitude. The time average of population imbalance is not zero, which indicates that Josephson oscillation is suppressed and self-trapping of population of particles occurs.

As we discussed earlier, under our chosen initial condition, previous studies of the dynamics of condensate in a double-well potential focused on the time evolution of the population imbalance between two wells. The standard two-mode theory only considers the two localized ground states. In this case, only the amplitudes of these two modes vary in time, which leads to the transfer of the particle population without altering the overall condensate density profile.  However, our numerical results show that the density profile of the condensate changes over time, indicating that more modes are involved under this condition. The simple two-mode theory can no longer be applied to this situation. We need to consider more energy levels. Nevertheless, the core assumption of the two-mode theory, namely that only the ground state mode and the first excited mode are populated at very low temperature, still holds in essence.
\section{\label{sec:level 4}Summary}
In summary, we have proposed a multi-mode model to investigate the dynamics of a BEC in a double-well potential. Under our approximation, the system can be effectively described by two independent two-mode models, one for the two local ground states and another for the two local first excited states. Numerical simulations show that a local collective excitation, i.e., the dipole oscillation, can undergo Josephson oscillations between the two wells and become self-trapping when the nonlinear interaction is sufficiently strong, as well as the similar dynamics of the population imbalance of condensate in double-well potential described by the standard two-mode model. Notably, these phenomena can be explained by our multi-mode model.

\end{document}